\documentclass[fleqn,twoside]{article}
\usepackage{amsmath}  
\usepackage{epsf}

\usepackage{espcrc2a}

\title{Testing Dimensional Reduction in SU(2) Gauge Theory}

\author{Slavo Kratochvila\address[ETHZ]{Institute for Theoretical Physics, ETH Z\"{u}rich, CH-8093 Z\"{u}rich, Switzerland}\thanks{Talk presented by S. Kratochvila} and
        Philippe de
        Forcrand\addressmark[ETHZ]\address[CERN]{Theory Division, CERN, CH-1211 Geneva 23, Switzerland}}

\begin{document}

\begin{abstract}

At high temperature, every $(d+1)$-dimensional theory can be
reformulated as an effective theory in $d$ dimensions. We
test the numerical accuracy of this Dimensional Reduction for
(3+1)-dimensional $SU(2)$ by comparing perturbatively determined
effective couplings with lattice results as the temperature is
progressively lowered. We observe an increasing disagreement
between numerical and perturbative values from $T=4 T_c$ downwards,
which may however be due to somewhat different
implementations of dimensional reduction in the two cases.

\end{abstract}

\maketitle

\section{INTRODUCTION}

One of the major thrusts of nuclear physics in the next decade
will be the effort to study the quark-gluon plasma. At the high
temperatures reached, one can consider applying Dimensional
Reduction. To test the accuracy of Dimensional Reduction, a purely
gluonic system provides a simple, but relevant benchmark.
Therefore, here we study $SU(2)$ gauge theory at finite
temperature. Up to now, only gauge invariant correlation functions
of colourless states have been
measured. 
This approach gives no information about the effective action.
Here, we measure directly the effective coupling constants. The
analytics of Dimensional Reduction gives us perturbative
expressions for these coupling constants. A numerical,
non-perturbative determination allows us to monitor the two ways
in which the perturbative calculation can fail: $(i)$ As the
temperature is lowered, the couplings increasingly depart from
their perturbative values. $(ii)$ The effective action becomes
less local - couplings of higher-dimension operators, e.g.
$A_0^6$, completely
neglected in the perturbative approach, become sizeable.\\
 The non-perturbatively determined
coupling constants allow us, with certainty in principle, to
extend the validity of Dimensional Reduction below temperatures
achieved perturbatively.
 In that sense, we attempt
a non-perturbative improvement of the dimensionally reduced
action.

\section{DIMENSIONAL REDUCTION}

The leading infrared behavior of $4d$-$SU(2)$ gauge theory at high
temperatures is governed by its static (zero Matsubara frequency)
sector, obtained by integrating out its non-static modes to leave
behind an effective three-dimensional theory. The non-static modes
are suppressed in the infrared.
At high temperature, $g(T)$ becomes small
due to asymptotic freedom. A perturbative treatment of the
non-static modes is justified. This is called Dimensional Reduction.\\
The Euclidean action is \begin{equation}\label{eq:$4d$action}
S^{4d,E}= \int \text{d}^{3} {\bf  x} \int_{0}^{\beta}\text{d}\tau
\frac{1}{4} F_{\mu\nu}^{a}F^{\mu\nu,a}. \end{equation} Expanding
the $A_\mu$-fields into Fourier series in the temporal direction,
keeping only the static modes and integrating out all the others,
we end up with a $3d$-$SU(2)$ gauge theory minimally coupled to a
self-interacting adjoint scalar field \cite{Kajantie:1996dw}:
\begin{align}
\label{eq:$3d$action}
 S^{3d,eff}=& \int \text{d}^{3} {\bf  x} \{
\frac{1}{4} {F}_{i j,0}^{a}{F}_0^{i j,a} +
\mathrm{Tr}[{D}_j,{A}_{0,0}^a][{D}_j,{A}_{0,0}^a]
\notag \\
& + m_D^2 \mathrm{Tr} ({A}_{0,0}^a)^2 + \lambda_A (\mathrm{Tr}
({A}_{0,0}^a)^2)^2 \}
\end{align}
where the first term is the $3d$-$SU(2)$ gauge theory and the
second term is the kinetic term of the static adjoint Higgs Field
${A}_{0,0}$. The coefficients of this $3d$ effective action, the
gauge coupling $g_3$, the Debye mass $m_D$ and the quartic
coupling $\lambda_A$, can be determined from the perturbative
expansion, or numerically as follows.

\section{CANONICAL DEMON METHOD}
Any action can be parameterized by \begin{equation} S=-\sum_\alpha
\beta_\alpha S_\alpha, \end{equation} where $S_\alpha$'s are
interaction terms, and  the $\beta_\alpha$'s are unknown effective
coupling constants.
 The lattice action corresponding to the continuum theory (\ref{eq:$3d$action})
  is after some reformulation \cite{Kajantie:1997tt}:
\begin{align}\label{latticeaction}
 S&=\beta_G \sum_{x, i<j}\left[ 1- \frac{1}{2}\mathrm{Tr} U_i(x)
U_j(x+\hat{i}) U_i^\dagger(x+\hat{j}) U_j^\dagger(x) \right] \notag \\
 &+\beta_A \sum_{x, i} \frac{1}{2}\mathrm{Tr} \tilde{A}_0(x) U_i(x) \tilde{A}_0(x+\hat{i}) U_i^\dagger(x) \notag \\
&+ \sum_{x}\left[-\beta_2 \frac{1}{2}\mathrm{Tr}\tilde{A}_0^2 +
\beta_4 ( \frac{1}{2}\mathrm{Tr}\tilde{A}_0^2)^2 \right]
\end{align}
where $a$ is the lattice spacing, $\beta_G=4/(a g_3^2)$,
$\beta_2=\left[3+(m_D a)^2/2\right]\beta_A$ and $\beta_4=a
\lambda_A \beta_A^2/4$. $\beta_A$ is arbitrary under a rescaling
of $\tilde{A}_0 \propto a A_{0,0_k} \sigma_k $. In addition we
introduce an auxiliary system, called demon system, given by the
action \begin{equation} S_D=-\sum_{\alpha=G,A,2,4} \beta_\alpha
d_\alpha, \end{equation} here the $\beta_\alpha$'s are the same as
in $S$, and the $d_\alpha$'s are the "demon energies", constrained
to lie in the interval $[-d_{max},d_{max}]$. The total partition
function factorizes into the one of the original system and that
of the single demons. We can compute the distribution of each
demon energy $d_\alpha$. Solving this equation with respect to
$\beta_\alpha$ allows us to determine the effective coupling
constants: \begin{equation} \langle d_\alpha
\rangle=\frac{1}{Z}\int d_\alpha e^{-\beta_\alpha d_\alpha}
\text{d} d_\alpha
=\frac{1}{\beta_\alpha}-\frac{d_{max}}{\text{tanh}(\beta_\alpha
d_{max} )}. \label{eq:average} \end{equation} We follow closely
the method of \cite{Hasenbusch:1994ne}. First, we generate a
statistically independent configuration C with a canonical update
of the $4d$-$SU(2)$-System only.  On this given configuration C,
we perform one microcanonical update of the
($4d$-$SU(2)$+Demon)-System, as follows: The change of a link,
under multiplication by a randomly generated matrix close to the
identity, causes a shift of the effective energy of the lattice,
$\Delta S_\alpha$, which must be compensated with the demon
energies $d_\alpha$. The trial is accepted if all the new demon
energies remain in the allowed region. After visiting all links,
the demons "move" to another gauge configuration C' while keeping
their energies. Because our ansatz for the $3d$-reduced action is
correct in the presence of small fluctuations only, we restrict
the microcanonical update to a small step size, and perform only
one microcanonical sweep on each gauge configuration.

\section{RESULTS}

The $\beta_{G,A,2,4}$ are related to the physical coupling
constants $g_3(\mu)$,$m_D^2(\mu)$ and $\lambda_A(\mu)$, as defined
after (\ref{latticeaction}). For $m_D^2(\mu)$ we must add
counterterms, calculated in \cite{Laine:1995ag}, to remove
UV-divergencies and compare with continuum $\overline{MS}$-scheme
theoretical calculations.
We measure the effective couplings as a function of temperature
$T/T_c$. To monitor effects of the lattice discretization, we
study, at each temperature, several systems of identical physical
size and increasingly fine lattice spacing ($N_t=2, 3, 4$ and
$5$).

\subsection{$g^2_3(\mu)$-results}

As shown in Fig.~1, we observe a clear, large disagreement between
the measurements and the perturbative prediction. The measured
values of the coupling appear to depend only mildly on the lattice
spacing. Unfortunately, the disagreement with the perturbative
prediction worsens as the continuum limit is approached. The
origin of the problem, we believe, lies in our lattice-based
approach. The perturbative calculation is performed in the
continuum theory, and the Green's functions are computed for small
spatial momenta $|\vec{p}| \ll T$. In contrast, the spatial
momenta on our lattices are cut-off by the lattice spacing, not by
the temperature. As we reduce the lattice spacing to approach the
continuum limit, we allow for larger momenta and depart more and
more from the infrared, perturbative result.

As a first attempt to mimic the effect of our lattice approach, we
tried to replace $\Lambda_{\overline{MS}} \rightarrow
\Lambda_{\overline{MS}} * f$ in the continuum perturbative
predictions. As shown in Fig.~1, a factor $f \sim 300$ is
necessary to reach reasonable agreement with our data, to be
compared with the value $19.82$ relating the $\overline{MS}$ and
the lattice renormalization schemes \cite{HH80}. In Figs.2 and 3,
the dashed line shows the effect of this rescaling $f=300$
on the other couplings. \\
\centerline{\epsfxsize=5.5cm\epsffile{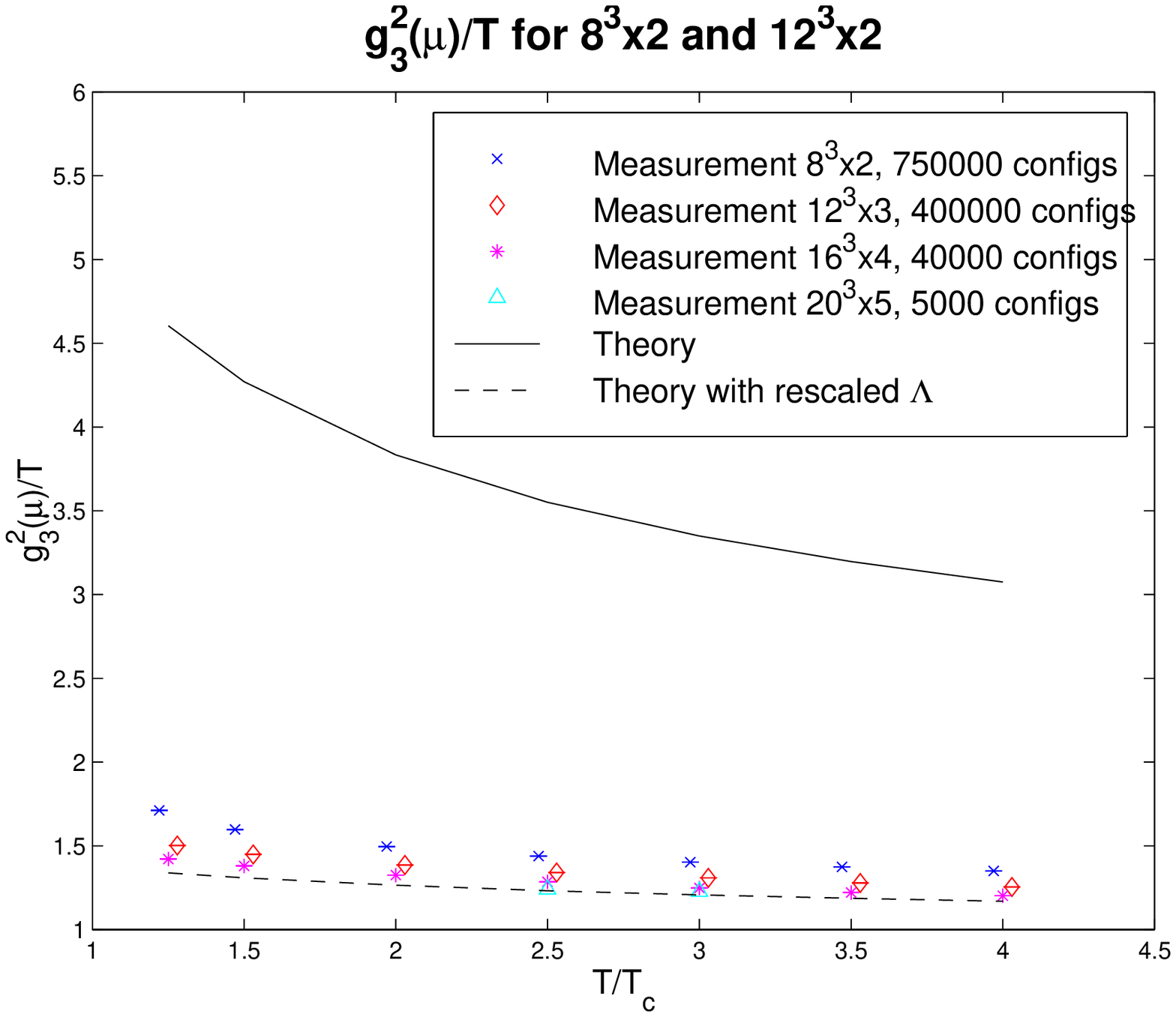}} \\
Figure 1. $g_3^2(\mu)/T$ vs $T/T_c$. Measurements and perturbative
predictions (solid line) clearly disagree. Rescaling
$\Lambda_{\overline{MS}}$ by a factor 300 restores agreement
(dashed line).

\subsection{$m_D^2(\mu)$-results}

$m_D^2(\mu)$ stays somewhat below the theoretical predictions for
all temperatures studied (up to $4 T_c$). Because the Debye mass
is only obtained after subtraction of UV-divergent counterterms,
statistical errors quickly increase as the continuum limit is
approached. There is a tendency for better agreement with
perturbation theory at smaller $a$. As explained above, agreement
should not be expected anyway since we keep larger spatial momenta
than in the continuum approach. Nevertheless, the disagreement for
$m_D^2$ is much smaller than for $g_3^2$, and could be accounted
for by a milder rescaling of the $\Lambda$ parameter.

\centerline{\epsfxsize=5.5cm\epsffile{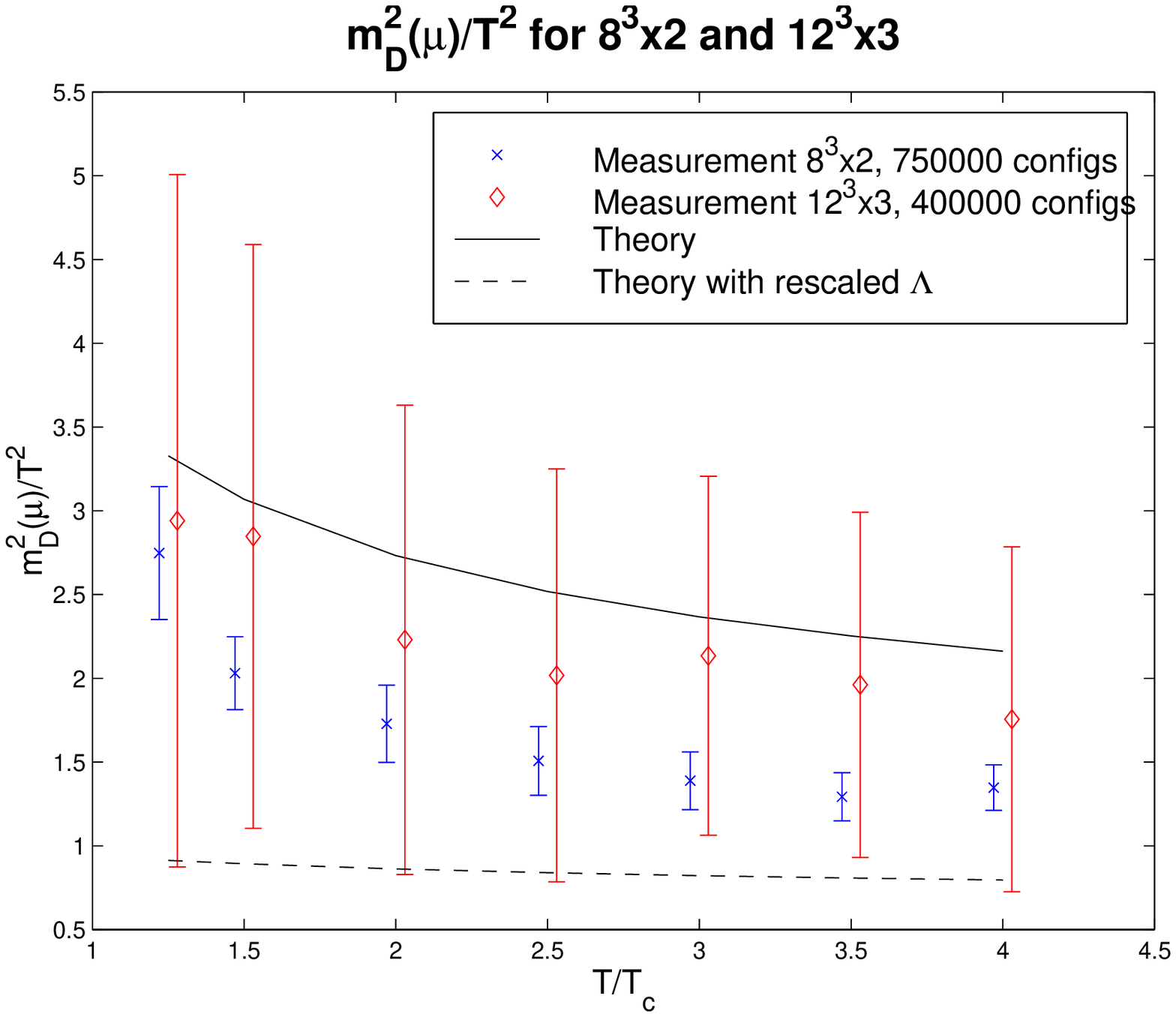}}
Figure 2. $m_D^2(\mu)/T^2$ vs $T/T_c$.
\subsection{$\lambda_A(\mu)$-results}

Like the other couplings, $\lambda_A(\mu)$ stays below the
perturbative
prediction at all temperatures. Scaling violations seem reasonably small.\\
\centerline{\epsfxsize=5.5cm\epsffile{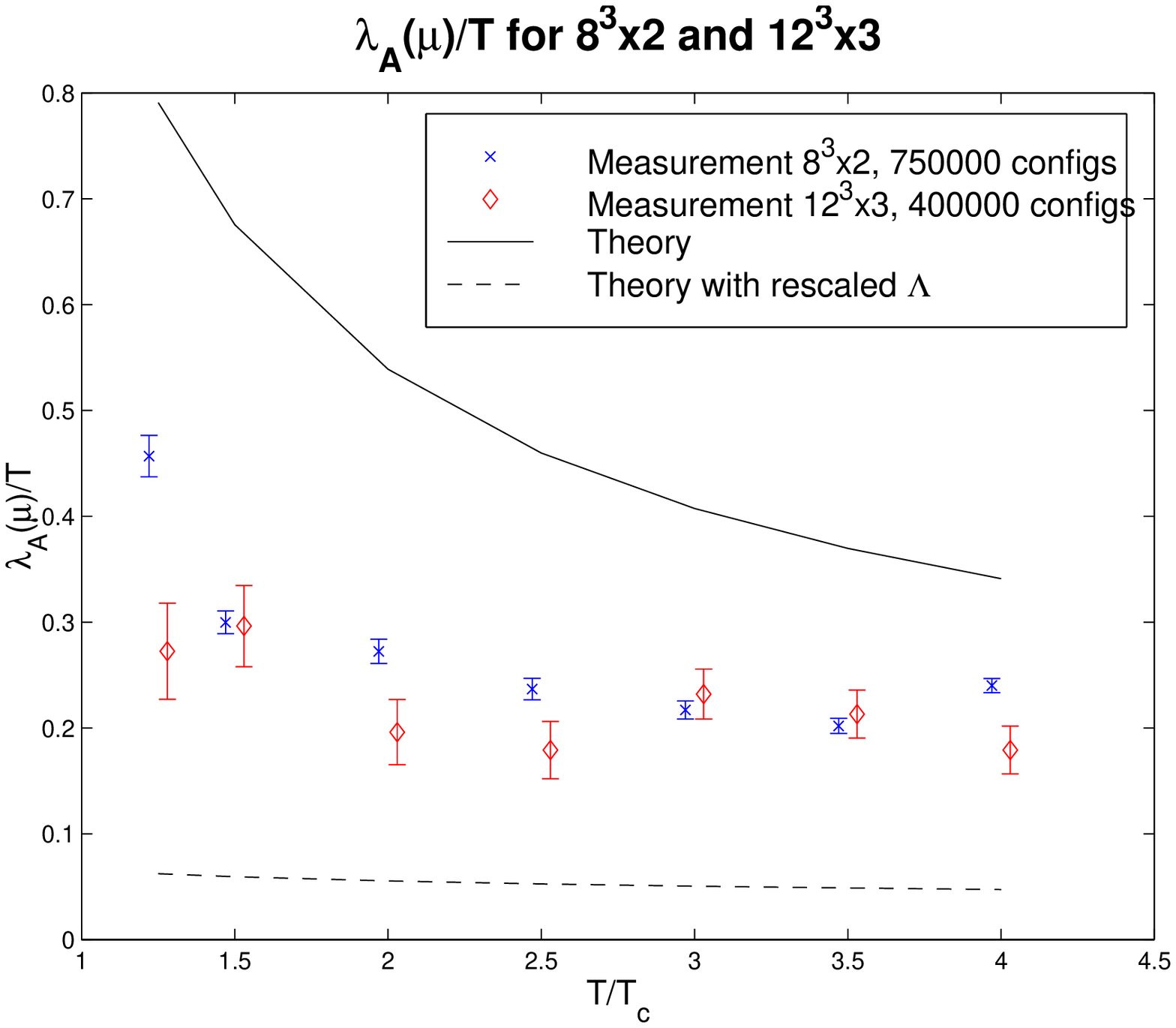}}\\
Figure 3. $\lambda_A(\mu)/T$ vs $T/T_c$.
\section{CONCLUSION}

In perturbative Dimensional Reduction, one considers only two and
four-point functions, to one- or at most two-loop level
\cite{Kajantie:1997tt}, and neglects higher-dimension operators.
Then, Dimensional Reduction appears to be valid down to $\sim 2
T_c$ \cite{Philipsen:2001et}.  Numerically determined effective
coupling constants allow a non-perturbative improvement of the
theory. Higher-dimension operators, with couplings determined by
additional demons, are easily included. We have added an
$A_0^6$-term in  the effective action, but the effect on the other
couplings is small. Our results still disagree with continuum
perturbative predictions even at $4 T_c$. We intend to compare
them with the perturbative
results of \cite{Petersson}, obtained directly on the lattice. \\



We are grateful to Mikko Laine for many helpful discussions and
advice.

\end{document}